\documentclass[%
 reprint,
superscriptaddress,
 amsmath,amssymb,
 pre,
]{revtex4-2}

\usepackage{graphicx}
\usepackage{dcolumn}
\usepackage{bm}

\usepackage{xcolor}

\newcommand{\nvec}[0]{{\bm n}}
\newcommand{\uvec}[0]{{\bm u}}

\begin{document}


\title{Liquid crystal torons in Poiseuille-like flows}
\author{Guilherme N. C. Amaral}
\affiliation{Centro de Física Teórica e Computacional, Faculdade de Ciências, Universidade de Lisboa, 1749-016 Lisboa, Portugal.}
 \affiliation{Departamento de Física, Faculdade de Ciências, Universidade de Lisboa, P-1749-016 Lisboa, Portugal.}
\author{Hanqing Zhao}
\affiliation{Department of Physics and Soft Materials Research Center, University of Colorado Boulder, CO 80309, USA.}
\author{Mahmoud Sedahmed}
\affiliation{Independent researcher, Egypt.}
\author{Tomás Campante}
\affiliation{Centro de Física Teórica e Computacional, Faculdade de Ciências, Universidade de Lisboa, 1749-016 Lisboa, Portugal.}
 \affiliation{Departamento de Física, Faculdade de Ciências, Universidade de Lisboa, P-1749-016 Lisboa, Portugal.}
\author{Ivan I. Smalyukh}
\email{ivan.smalyukh@colorado.edu}
\affiliation{Department of Physics and Soft Materials Research Center, University of Colorado Boulder, CO 80309, USA.}
\affiliation{Department of Electrical, Computer, and Energy Engineering and Materials Science and Engineering Program, University of Colorado, Boulder, CO 80309.}
\affiliation{Renewable and Sustainable Energy Institute, National Renewable Energy Laboratory and University of Colorado, Boulder, CO 80309, USA.}
\affiliation{International Institute for Sustainability with Knotted Chiral Meta Matter, Hiroshima University, Higashihiroshima 739-8511, Japan.}
\author{Mykola Tasinkevych}
\email{mykola.tasinkevych@ntu.ac.uk}
  \affiliation{Centro de Física Teórica e Computacional, Faculdade de Ciências, Universidade de Lisboa, 1749-016 Lisboa, Portugal.}
 \affiliation{Departamento de Física, Faculdade de Ciências,
Universidade de Lisboa, P-1749-016 Lisboa, Portugal.}
\affiliation{SOFT Group, School of Science and Technology, Nottingham Trent University, Clifton Lane, Nottingham NG11~8NS, United Kingdom.}
\affiliation{International Institute for Sustainability with Knotted Chiral Meta Matter, Hiroshima University, Higashihiroshima 739-8511, Japan.}
\author{Margarida M. Telo da Gama}%
  \affiliation{Centro de Física Teórica e Computacional, Faculdade de Ciências, Universidade de Lisboa, 1749-016 Lisboa, Portugal.}
 \affiliation{Departamento de Física, Faculdade de Ciências,
Universidade de Lisboa, P-1749-016 Lisboa, Portugal.}
\author{Rodrigo C. V. Coelho}
\affiliation{Centro de Física Teórica e Computacional, Faculdade de Ciências, Universidade de Lisboa, 1749-016 Lisboa, Portugal.}
 \affiliation{Departamento de Física, Faculdade de Ciências, Universidade de Lisboa, P-1749-016 Lisboa, Portugal.}
\email[]{rcvcoelho@fc.ul.pt}


\date{\today}

\begin{abstract}
Three-dimensional (3D) simulations of the structure of liquid crystal (LC) torons, topologically protected distortions of the LC director field, under material flows are rare but essential in microfluidic applications. Here, we show that torons adopt a steady-state configuration at low flow velocity before disintegrating at higher velocities, in line with experimental results. Furthermore, we show that under partial slip conditions at the boundaries, the flow induces a reversible elongation of the torons, also consistent with the experimental observations. These results are in contrast with previous simulation results for 2D skyrmions under similar flow conditions, highlighting the need for a 3D description of this LC soliton in relation to its coupling to the material flow. These findings pave the way for future studies of other topological solitons, like hopfions and heliknotons, in flowing soft matter systems.
\end{abstract}

\maketitle


\makeatletter
\newcommand{\manuallabel}[2]{\def\@currentlabel{#2}\label{#1}}
\makeatother


\section{Introduction}

Liquid crystals (LCs) are a class of materials that exhibit both fluidity and anisotropy, allowing for a diverse range of unique applications. The intrinsic ability of the LC director field to respond to external stimuli such as electric and magnetic fields, as well as the ease with which boundary conditions can distort it, underpins the functionality of modern display technologies~\cite{Chen2018ER}. Among the various phases of LCs, chiral nematics are particularly intriguing due to their ability to host a variety of stable spatially localized topological solitons, including torons, hopfions and skyrmions, which have garnered significant attention for their potential in both fundamental research and technological applications~\cite{Smalyukh2010, Ackerman2014, Ackerman2017}. We will refer as skyrmions to the 2D version of 3D torons, or their midplane cuts.

Skyrmions in LCs, much like their counterparts in magnetic systems, are localized, non-singular configurations of the director field that cannot be continuously transformed into a uniform state~\cite{Skyrme1962, Tai2018}. These solitons, whether in 2D or 3D, exhibit rich structural behaviours and conformational transitions, making them a promising platform for exploring topological phenomena in soft matter systems~\cite{Ackerman2017, Fukuda2011, Posnjak2016, Ackerman2017b, guo:2016, tai:2019,  Zhao2023, sohn2018}. The study of topological solitons in LCs is not only of academic interest but has practical applications, as these structures can be manipulated using external fields and exhibit particle-like behaviour, including interactions that can be tuned from attractive to repulsive~\cite{Sohn2019, Sohn2020, D3SM01422C, Alvim2024}. 

Despite extensive research on LC solitons, the impact of externally imposed material flows on their behaviour remains poorly understood. Previous studies have predominantly focused on the dynamics induced by electric fields, overlooking the role of hydrodynamic interactions with a few exceptions~\cite{coelho2024halltransportliquidcrystal, Coelho_2021}. However, in applications such as microfluidics and lab-on-a-chip devices~\cite{C6LC00387G, DKHAR2023115120}, solitons will interact with material flows, requiring a deep understanding of their behaviour in such environments.

In previous work~\cite{PhysRevResearch.5.033210}, we used experiments and 2D simulations to investigate the behaviour of skyrmions in Poiseuille-like flows. We observed a persistent skyrmion elongation with a characteristic time given by the ratio of the cholesteric pitch and the average flow velocity. However, the limitations of 2D simulations became apparent when compared to 3D experimental systems. Notably, while skyrmions in 2D simulations remained elongated after turning off the flow, in experiments, they relaxed back to their original quasispherical shapes. Additionally, the 2D simulations predicted elongation at any flow velocity, whereas in experiments, skyrmions only elongated above a certain velocity threshold.

In this paper, we address these discrepancies by extending the simulations to three dimensions. We found that the 3D simulations successfully describe the qualitative behaviour observed in the experiments. Specifically, we show that torons acquire a steady-state shape at low flow velocity and are destroyed at high velocity, consistent with experimental observations reported here. Moreover, by imposing partial slip at the boundaries, we recover the elongation behaviour observed in the experiments at higher velocities~\cite{PhysRevResearch.5.033210}, possible due to the effect of a high-frequency electric field that stabilizes the skyrmion. When the external flow is turned off, the elongated torons in the 3D simulations revert to their original shape, in line with the experimental results.

\section{Results}

\subsection{Steady-state shape}

\begin{figure}[ht]
\includegraphics[width=\linewidth]{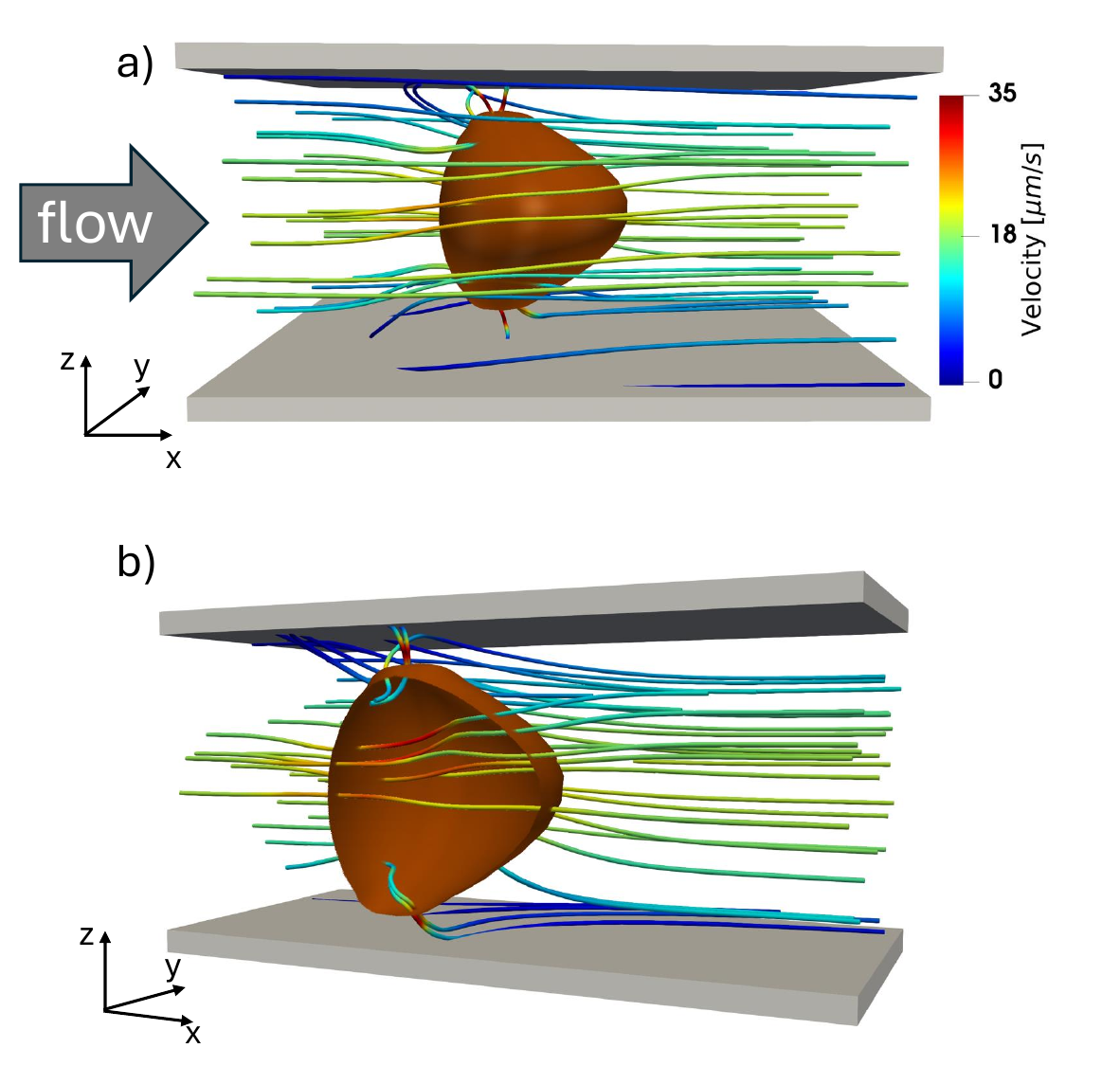}
\caption{Velocity field around the toron moving with velocity $v_{t}=19.5 \, \mu$m/s, corresponding to the Ericksen number $Er=2.8$ defined in the main text. (a) Full toron and (b) cut through the toron, to highlight the inner flow structure, depicted by the isosurface (in red) $n_z=1$, where $n_z$ is the $z$ component of the director field.}
\label{Fig:scheme}
\end{figure}
\begin{figure*}[ht]
\includegraphics[width=\textwidth]{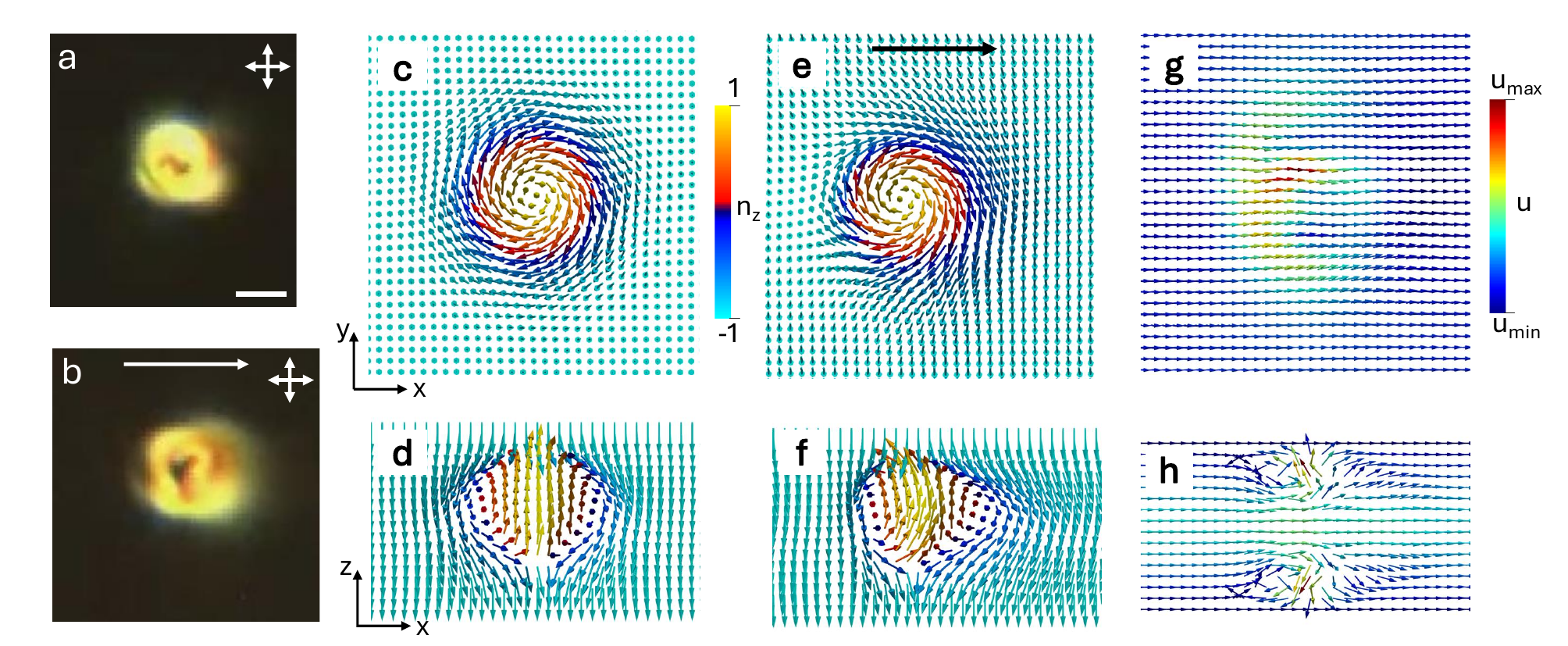}
\caption{Polarizing optical microscopy images of torons in zero flow (a), and moving at 17 $\mu$m/s ($Er=2$) under steady-state conditions (b); the white arrow gives the direction of the flow. The scale bar in (a) corresponds to $10 \,\mu$m. See video 1 in the Supplementary Material. (c) and (d) depict simulation results of the static toron structure (zero flow). Cross-sectional views of the director field are shown by the coloured arrows in the planes $z=L_Z/2$ (c) and $y=L_Y/2$ (d), with colour representing the $z$ component of the director field according to the legend. (e) and (f) simulation results of the configuration of a toron moving at $v_{t}=19.5 \, \mu$m/s ($Er=2.8$); the direction of the flow velocity is given by the black arrow in (e). Director field in the planes $z=L_Z/2$ (e) and $y=L_Y/2$ (f). (g) and (h) depicts the flow field in the planes $z=L_Z/2$ and $y=L_Y/2$, respectively, around a toron moving at $v_{t}=19.5 \, \mu$m/s. In (g), the red and blue represent $u_{\text{max}}=17$ and $u_{\text{min}}=8\, \mu$m/s respectively while, in (h), they represent $u_{\text{max}}=55$ and $u_{\text{min}}=0\, \mu$m/s.}
\label{Fig:shape}
\end{figure*}
We start by describing the behaviour of a stable toron in the steady-state regime, i.e., with a steady shape.
The simulation setup consists of a cholesteric liquid crystal between two parallel plates where we impose no-slip boundary conditions and rigid homeotropic anchoring. The toron is initialized using a director ansatz that imposes the correct director topology and this configuration is relaxed until it reaches convergence. An external body force (or acceleration) $g$ is applied to the liquid crystal that makes it flow at a constant average velocity in the steady-state. Figure~\ref{Fig:scheme} illustrates the setup with the toron contour indicated in red. The coupling between the flow and director fields leads to a noticeable change in the direction of the streamlines in the vicinity and inside the toron. It is useful to describe the motion in terms of the Ericksen number, which is defined as the ratio of the viscous and elastic forces, given by $Er = \frac{\alpha_4 v_{t} L_Z}{K_{11}}$, where $\alpha_4$ is one of the Leslie viscosity parameters, $v_{t}$ is the toron velocity, $L_Z$ is the separation between the plates, and $K_{11}$ is the splay elastic constant.

In the experiments at low Ericksen numbers (Fig.~\ref{Fig:shape}), torons move with a steady shape as depicted in Fig.~\ref{Fig:shape}(b). The toron shape is slightly distorted compared to the static one (Fig.~\ref{Fig:shape}(a)), but it is steady, which was not found in the 2D simulations where the skyrmions continuously elongated at any $Er$~\cite{PhysRevResearch.5.033210}. By contrast, the 3D simulations describe the experimentally observed steady-state toron distortions, as illustrated in Fig.~\ref{Fig:shape}(e) and (f). Interestingly, the toron becomes asymmetric slightly elongating in the flow direction. This distortion induced by the flow field is reversible and will be discussed in more detail in Sec.~\ref{relaxation-sec}. The flow field is also distorted in the inner region of the toron and at the hyperbolic hedgehog defects close to the confining plates as shown in Fig.~\ref{Fig:shape}(g) and (h). These are the regions with the largest director field gradients. Far from the toron, the velocity field is similar to that of a Poiseuille flow.

\begin{figure}[ht]
\includegraphics[width=\linewidth]{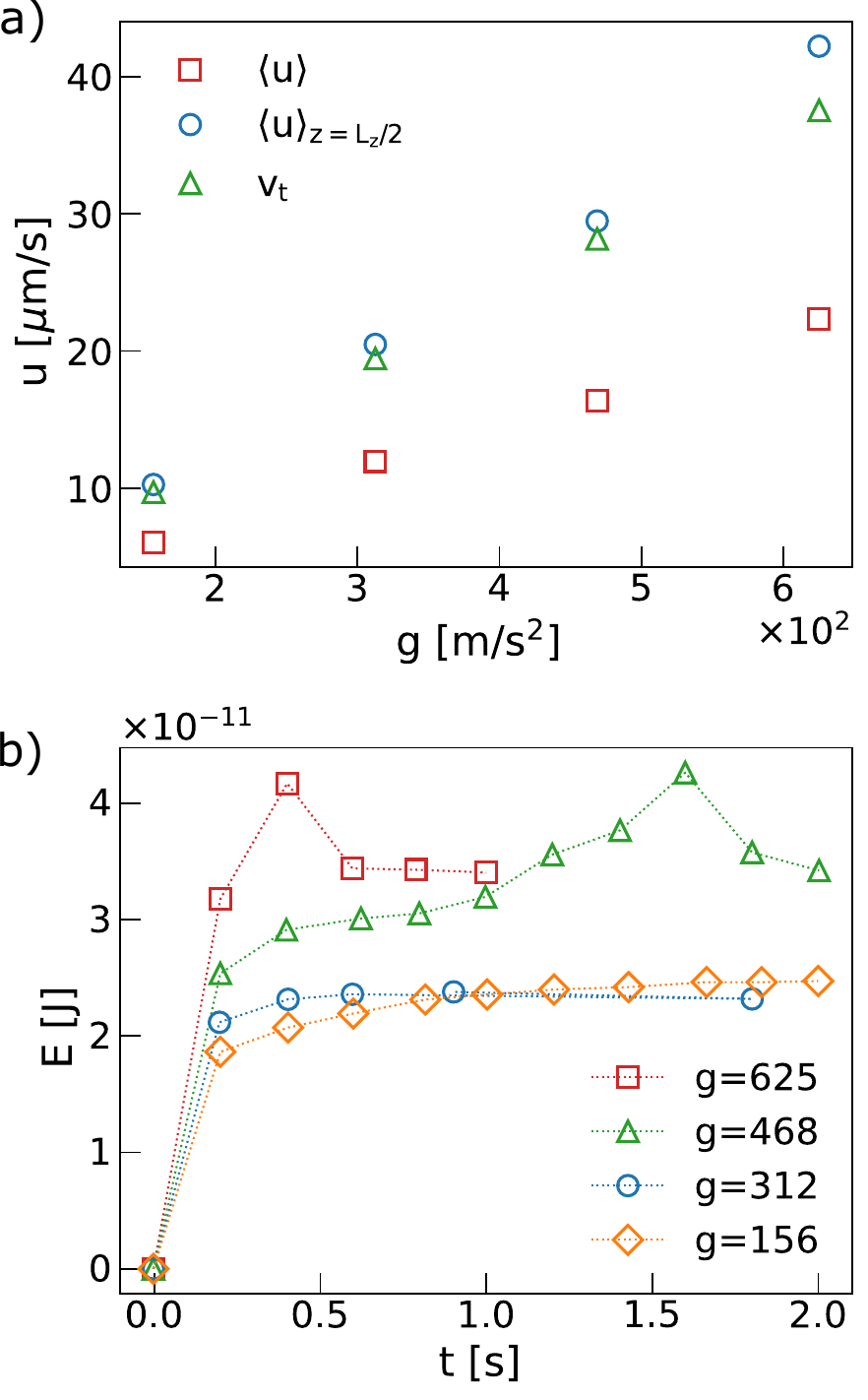}
\caption{(a) Comparison between the toron velocity $v_{t}$, the average flow velocity $\langle u \rangle$, and the flow velocity $\langle u \rangle_{z=L_Z/2}$ at the cell midplane $z=L_Z/2$. For the two largest forces, the toron is destroyed after some time and the measurement considers the toron velocity before that. The Ericksen numbers for the 4 simulations are, from low to high velocities: 1.4, 2.8, 4.0, and 5.3. (b) Time evolution of the Frank-Oseen free energy for four different external forces (legend, in m/s$^2$). }
\label{Fig:energy}
\end{figure}

The 3D flow field is intrinsically heterogeneous: it is maximal in the midplane at the centre of the cell and zero at the plates. The toron velocity $v_{t}$, defined by the velocity of the geometric centre of the region with $n_z>0$, is considerably larger than the average fluid velocity $\langle u\rangle$,  $v_{t}\approx 1.66\, \langle u \rangle$, but it is slightly smaller than the average velocity at the midplane $z=L_Z/2$ as indicated in Fig.~\ref{Fig:energy}(a) for different external forces $g$. The free energy $E$, Fig.~\ref{Fig:energy}(b), reaches a plateau at late times for low $Er$ (or forces), which indicates that the system has reached a steady-state. For larger $Er$, $E$ exhibits peaks, which correspond to the disintegration of the torons as will be discussed in the next section.

\subsection{Unstable toron}

\begin{figure*}[htb]
\includegraphics[width=0.8\textwidth]{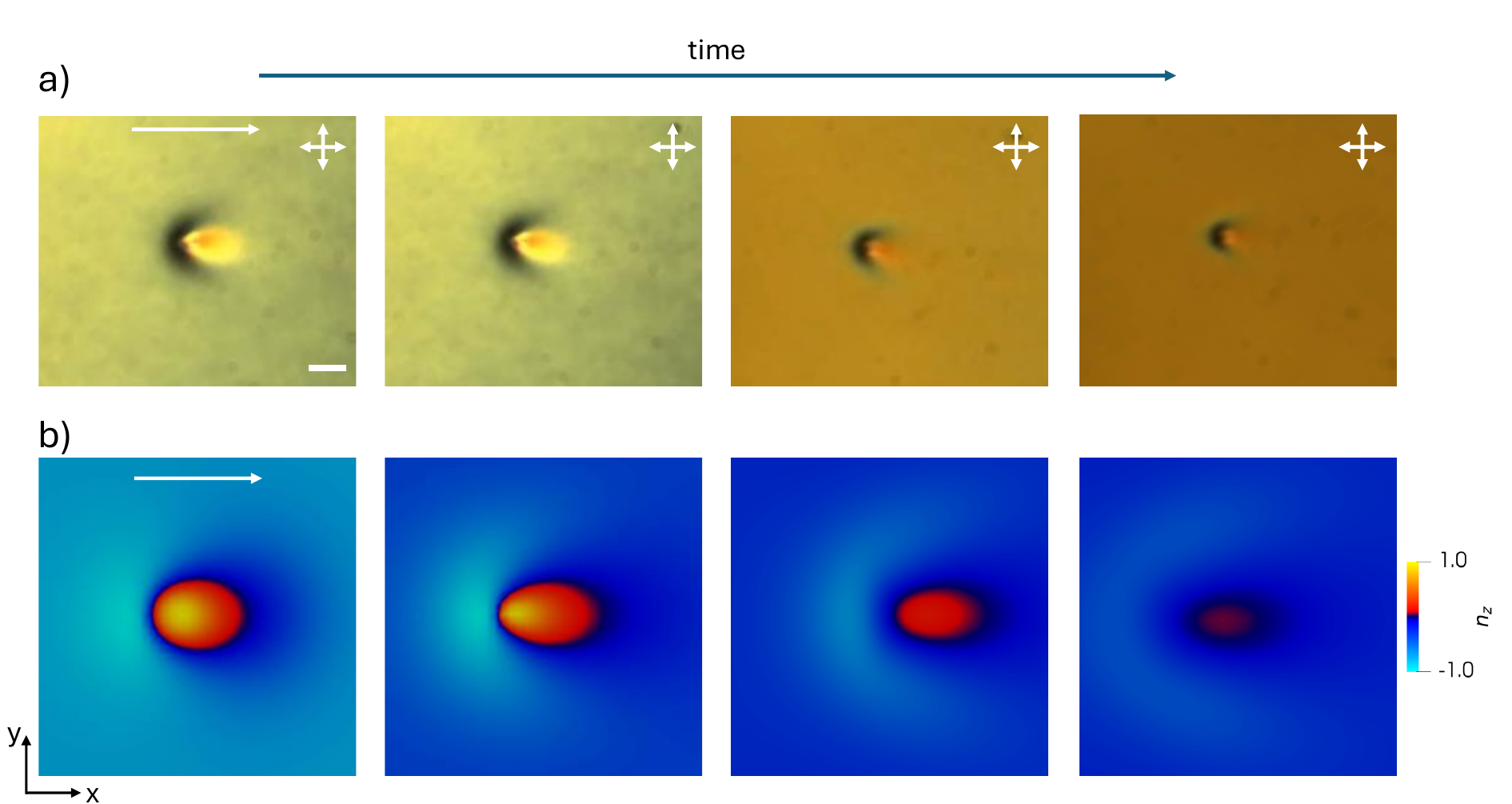}
\caption{Destruction of a flowing toron at high flow velocity ($Er$). (a) Experiment: POM images of the toron configurations moving with velocity 80 $\mu$m/s (Er=9.3). The white scale bar corresponds to $10\, \mu$m. See video 2 in the supplementary material. (b) Simulations with toron velocity $v_{t}=37.5\, \mu$m/s ($Er=5.3$). The white arrow indicates the direction of the flow.}
\label{Fig:destruction}
\end{figure*}

In our previous work, based on simulations in 2D~\cite{PhysRevResearch.5.033210}, the skyrmion was found to be stable even at very high velocities. This is due to the fact that the 2D flow is almost uniform (not parabolic as in 3D), which renders the skyrmion much more stable. Here, using 3D simulations, we found that the toron either moves with a constant shape as described in the previous section or becomes unstable above a certain velocity due to shear. Figure~\ref{Fig:destruction} depicts snapshots of the skyrmion configuration during the destruction process, which is qualitatively similar in experiments and simulations. In the simulations, torons become unstable for $v_{t}>23.9 \, \mu$m/s ($Er>3.4$). Initially, the toron becomes asymmetric and its configuration resembles that of a toron under an electric field in a chiral nematic system with negative dielectric anisotropy~\cite{Ackerman2017}. The far-field in the experimental polarized optical microscopy (POM) images becomes darker indicating that the director field aligns with the flow, and then the toron disappears. This event corresponds to the peaks observed in the free energy as a function of time, see the green and red curves in Fig.~\ref{Fig:energy}(b). After the toron destruction, the free energy stabilizes at a lower value, which is larger at higher flow velocities due to the larger inclination of the director field.

\subsection{Toron elongation}
\label{elongation-sec}

\begin{figure*}[htb]
\includegraphics[width=0.8\textwidth]{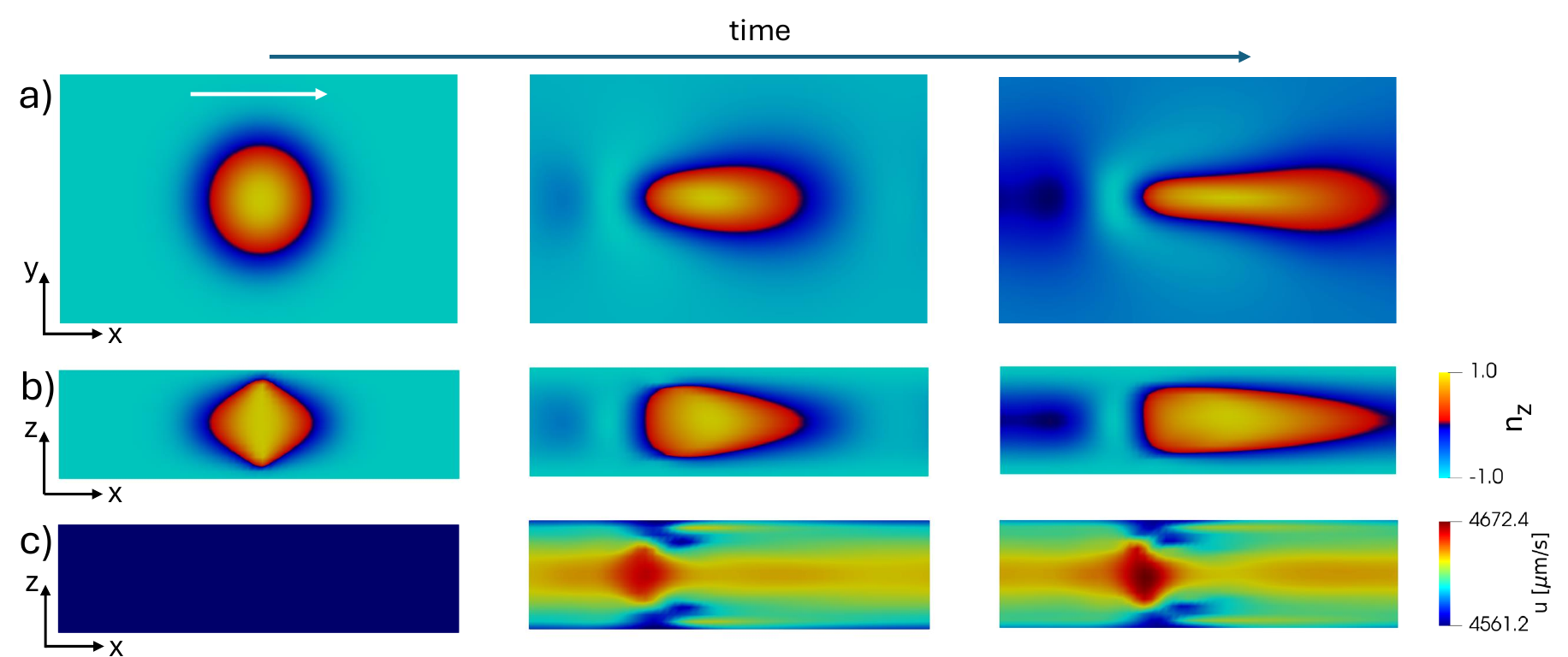}
\caption{Toron elongation with partial slip at the plates for driving forces $g=625$ m/s$^2$ and friction parameter $\chi=1.1\times 10^{6}$ s$^{-1}$. The director field in the planes (a) $z=L_z/2$ and (b) $y=L_y/2$ is shown by the color-coded $z$ component of the director field $n_z$. (c) Color-coded magnitude of the velocity field in the plane $y=L_y/2$. The white arrow indicates the direction of the flow. }
\label{Fig:elongation}
\end{figure*}

Here we propose an explanation for the skyrmion elongation observed at larger velocities in the experiments. Those higher velocities were obtained with a stabilizing high-frequency electric field with $f_{\text{stab}}=1000$ Hz and $U_{\text{stab}}=3.5$ V, as described in our previous work~\cite{PhysRevResearch.5.033210}. The ratio between the distance between plates and the pitch is $d/p=1.25$, for which, in zero field, the torons elongate and assume finger-like structures. In the 3D simulations, we did not observe elongation at any velocity when using the parameters given in Table~\ref{tab1}, strong anchoring and no-slip conditions at the confining plates. Either the toron moves with a constant shape or it becomes unstable. Thus, one possibility is that the liquid crystal is forced to flow so fast in the experiment that it partially detaches from the plates and the no-slip conditions no longer describe the experiments. To test this hypothesis, we impose partial slip conditions, which is done by assuming a friction force with the plates: $\mathbf{F}_d = - \chi \mathbf{u}$, where $\chi$ is the friction coefficient and $\mathbf{u}$ is the flow velocity of the fluid layers next to the plates. This force is applied only at the confining plates. The results of the previous sections can be recovered for large values of $\chi$. For smaller friction, the toron elongates, as depicted in Fig.~\ref{Fig:elongation}. In the midplane $z=L_Z/2$, the observed toron shapes are similar to those observed in the 2D simulations~\cite{PhysRevResearch.5.033210}. A perpendicular cross-sectional view in the plane $y=L_Y/2$ shows that the hyperbolic hedgehog defects at the confining plates remain behind while a ``belly'' in the central region of the toron extends continuously in the flow direction. The flow field is perturbed mostly close to the defects at the plates while being Poiseuile-like (albeit non-zero at the plates) far from the toron. 

\begin{figure}[htb]
\includegraphics[width=\linewidth]{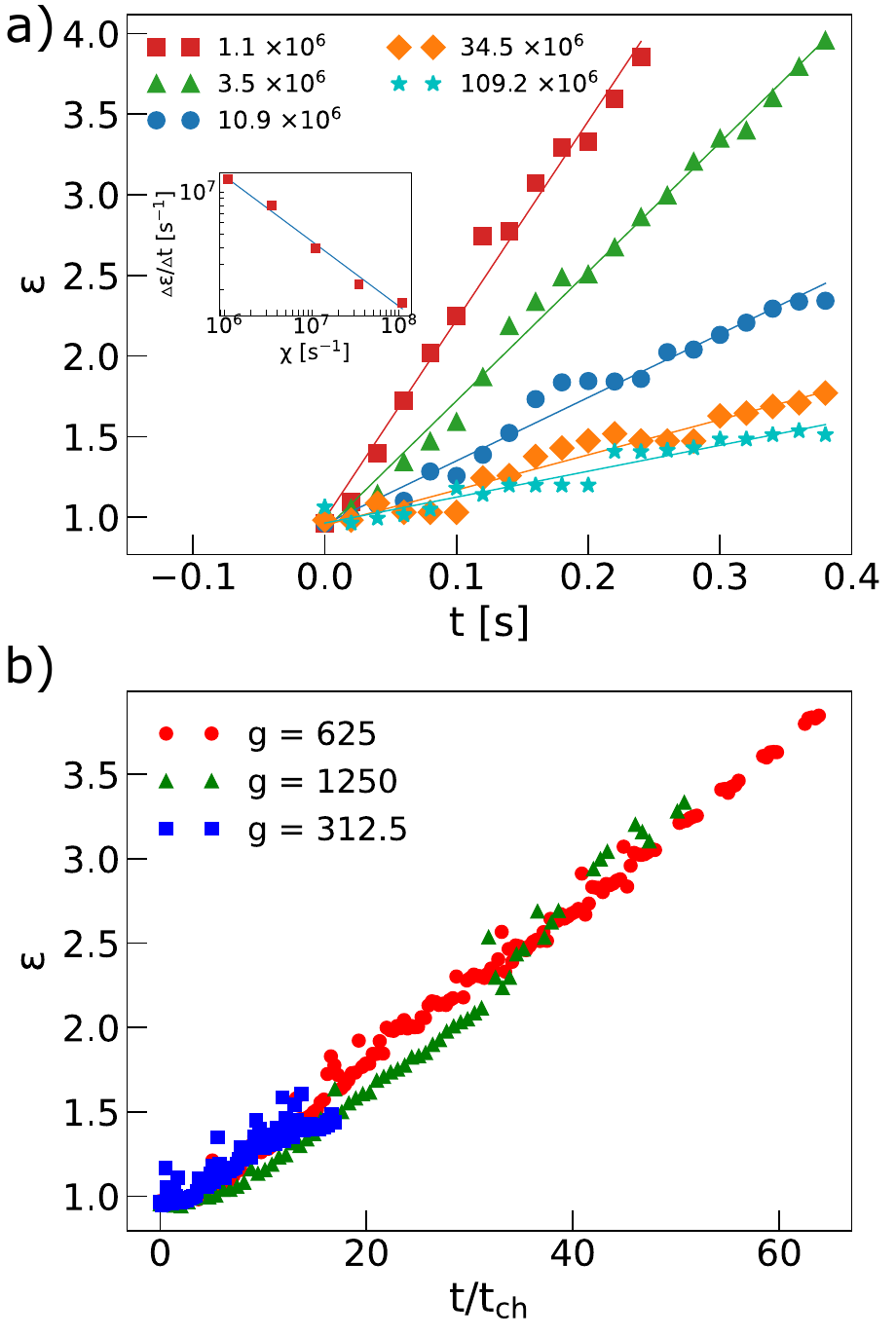}
\caption{Time evolution of the aspect ratio $\varepsilon$ of the skyrmions flowing between plates with partial slip conditions. (a) Results obtained at fixed force, $g=625$ m/s$^2$, and different friction coefficients (legend, in units of s$^{-1}$). The solid lines are linear fits which provide the elongation rates. The inset depicts the elongation rate against the friction coefficient in the log-log scale. The solid line is a power law fit with exponent $-0.47$. (b) Aspect ratio calculated at fixed friction, $\chi=3.5 \times 10^{-6}$ s$^{-1}$, and different forces $g$ (legend, in units of m/s$^2$). The time is in units of the characteristic time $t_{ch}=p/\langle u \rangle$. }
\label{Fig:friction}
\end{figure}

Figure~\ref{Fig:friction}(a) illustrates the time evolution of the skyrmion aspect ratio $\varepsilon$ at different friction coefficients $\chi$ keeping the external force density constant. The aspect ratio evolves linearly with time similar to the 2D results reported in ~\cite{PhysRevResearch.5.033210} with a rate that decreases with the friction until eventually reaching zero at infinite friction, in line with the results reported in the previous sections where the aspect ratio was constant. The persistent elongation observed under partial slip boundary conditions is consistent with that observed in the 2D simulations and in experiments with large velocities under stabilizing electric fields. Additionally, Fig.~\ref{Fig:friction}(b) indicates that the curves $\varepsilon(t)$ collapse when the time is rescaled by a characteristic time $t_{ch}=p/\langle u \rangle$, where $\langle u \rangle$ is the domain averaged flow velocity. This data collapse is analogous to that observed in the 2D simulations~\cite{PhysRevResearch.5.033210}, revealing that there is a characteristic time for the skyrmion elongation also in 3D systems.

\subsection{Relaxation without external flow}
\label{relaxation-sec}

One limitation of the 2D model was the description of the relaxation of the skyrmion shape after stopping the external flow. In 2D, due to the assumption of translational invariance perpendicular to the simulation plane, the skyrmion did not fully relax to the original circular shape but took an elongated symmetric shape after turning off the flow. This was inconsistent with the experimental observations, where the skyrmion returned to its original shape. 

\begin{figure*}[htb]
\includegraphics[width=\textwidth]{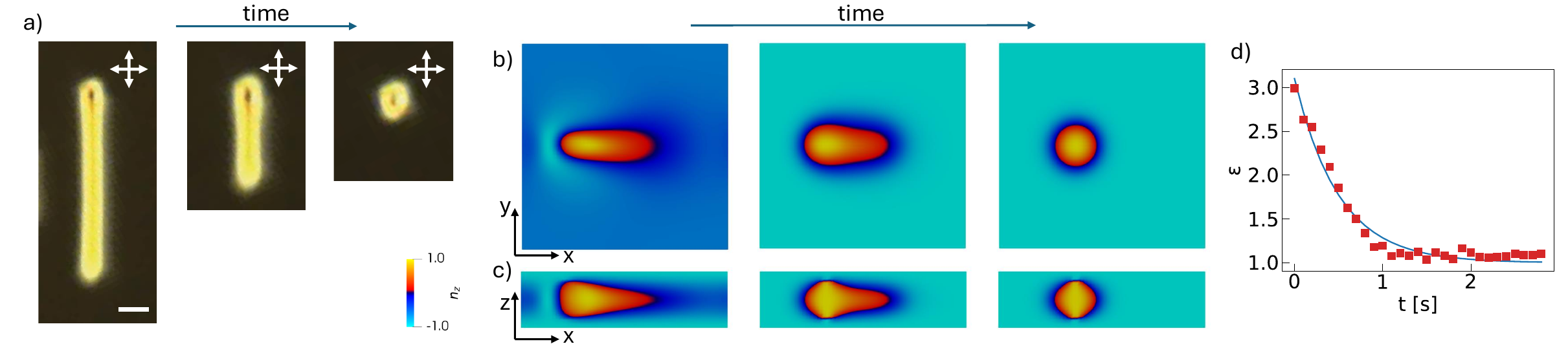}
\caption{Relaxation of an elongated toron after turning off the flow. (a) Experimental POM images where the scale bar corresponds to $10 \,\mu$m. See video 3 in the Supplementary Material. (b) and (c) color coded $n_z$ obtained from the simulations at the planes $z=L_Z/2$ and $y=L_Y/2$, respectively. (d) Time evolution of the aspect ratio for the relaxing toron. The solid line is an exponential fit of the form $\epsilon(t)=a\exp(-t/\tau_r)+1$, with $a=2.1$ and $\tau_r=0.5$ s.}
\label{Fig:relaxation}
\end{figure*}

We relaxed the system without external flow by using 3D simulations and starting from the elongated configuration obtained with partial slip conditions. As shown in Fig.~\ref{Fig:relaxation}, the results correctly describe the reversibility of the toron deformation. The time evolution of the aspect ratio indicates that it converges exponentially to its original quasispherical shape. Thus, the shape deformations of the toron driven by the external flow are, in general, reversible/elastic. Only when the configuration is invariant along the perpendicular direction, as in the 2D simulations, is the elongation persistent when the flow is turned off.

\section{Conclusions}

We have extended the theoretical study of liquid crystal torons in Poiseuille-like flows from two-dimensional (2D) to three-dimensional (3D) simulations, addressing key discrepancies observed between previous 2D simulations and the experimental results. The results of the 3D simulations reported above successfully describe the behaviour observed in the experiments, providing a deeper insight into the dynamics of torons under flow.

We found that in 3D, torons acquire a steady-state shape at low flow velocities, maintaining their structure without continuous elongation, and become unstable at higher velocities, leading to their eventual destruction. This behaviour is in line with the experimental observations. Moreover, by incorporating partial slip conditions at the boundaries, we reproduced the elongation of torons observed in experiments with a stabilizing electric field. We have not observed the elongation of torons when using no-slip conditions at the confining plates and strong anchoring. A possibility for further study is the detailed understanding of the anchoring strength as well as the effect of stabilizing electric fields.

The 3D modelling also successfully describes the toron shape's reversibility after turning off the external flow. By contrast to 2D simulations, where the torons remain elongated after turning off the flow, the 3D simulations reveal that torons return to their original shape, consistent with the experimental results. This highlights the importance of considering the full 3D nature of the system when studying the toron dynamics under external flows.

Our results bridge the gap between simulations and experiments, providing a rather accurate representation of the toron dynamics in 3D flow environments. This work advances our understanding of topological solitons in liquid crystals and paves the way for future research into the complex interplay between topological structures and hydrodynamic forces in soft matter systems, which may lead to new applications in microfluidic devices and advanced materials.

\section{Methods}



\subsection{Sample preparation and experimental methods}

Chiral liquid crystals are prepared by mixing 4-Cyano-4'-pentylbiphenyl (5CB, EM Chemicals) with a left-handed chiral additive, cholesterol pelargonate (Sigma-Aldrich). The pitch ($p$) of the resulting chiral LCs is determined by the weight fraction of the added chiral dopant, calculated using $C_{dopant}=1/(h_{htp})$, where the helical twisting power $h_{htp}$ is $6.25 \,\mu$m$^{-1}$ for cholesterol pelargonate.

The sample cells are constructed from glass slides treated with polyimide SE5661 (Nissan Chemicals) to achieve strong perpendicular (homeotropic) boundary conditions. The polyimide is applied by spin-coating at 2700 rpm for 30 s, followed by baking (5 min at 90 $^\circ$C and then 1 h at 180$^\circ$C).

The LC cell gap thickness is set to 10 $\mu$m using silica spheres as spacers, with a cell gap-to-pitch ratio $d/p = 0.85$. Under these conditions, the skyrmion is stable without the need for a stabilizing electric field. The lateral sides of the LC cell are sealed with epoxy, leaving two entrances. A custom connector created using a Formlabs Form 2 resin 3D printer connects one entrance to a tube, and the other end of the tube connects to a syringe. By manipulating the syringe plunger, the LC inside the cell channel is moved forward or backwards. 

A ytterbium-doped fiber laser (YLR-10-1064, IPG Photonics, operating at 1064 nm) is used to generate torons. The LC is locally melted using laser power around 30 mW, and upon switching off the laser tweezers, torons form spontaneously as the LC cools and quenches back. As the LC flows, the torons are carried along by the flow.

Polarized optical microscopy images are captured using a multi-modal imaging setup built around an IX-81 Olympus inverted microscope and charge-coupled device cameras (Grasshopper, Point Grey Research). Olympus objectives with 20x and 10x magnification and numerical apertures of NA=0.4 and 0.1 are employed. ImageJ software (NIH freeware) is used to analyze the speed, length, and trajectories of the torons.

\subsection{Simulations}

The dynamics of the liquid crystal (LC) director field are described by the Ericksen-Leslie model~\cite{Ericksen1962, doi:10.1098/rspa.1968.0195, stewart2019static}. This model consists of two sets of equations: one for the material flow and another for the director field. These equations are particularly adequate to describe the behaviour of LCs in the nematic or cholesteric phases.

For the velocity field, we use the Navier-Stokes equation along with the continuity equation:
\begin{eqnarray}
&&\rho \partial_t u_\alpha + \rho u_\beta \partial_\beta u_\alpha = \partial_\beta \left[ - P \delta_{\alpha\beta} + \sigma_{\alpha\beta}^{v} + \sigma_{\alpha\beta}^{e} +\rho g \right] \label{NS-eq} \\
&& \partial_\alpha u_\alpha = 0\label{cont-eq},
\end{eqnarray}
where the viscous stress tensor is defined as:
\begin{eqnarray}
\sigma_{\alpha\beta}^{v} = && \alpha_1 n_\alpha n_\beta n_\mu n_\rho D_{\mu\rho} + \alpha_2 n_\beta N_\alpha + \alpha_3 n_\alpha N_\beta \\
&&+ \alpha_4 D_{\alpha\beta} + \alpha_5 n_\beta n_\mu D_{\mu\alpha} + \alpha_6 n_\alpha n_\mu D_{\mu\beta} .
\end{eqnarray}
In these equations, $\rho$ denotes the fluid density, $P$ is the pressure, $\uvec$ is the fluid velocity, $\nvec$ is the director field indicating the preferred alignment of the LC molecules, $g$ is the external acceleration that drives the fluid, and $\alpha_n$'s are the Leslie viscosities. The kinematic transport, which describes the influence of the macroscopic flow field on the microscopic structure, is given by:
\begin{eqnarray}
N_\beta = \partial_t n_\beta + u_\gamma \partial_\gamma n_\beta - W_{\beta \gamma} n_\gamma
\end{eqnarray}
while the shear rate and vorticity tensors are defined as:
\begin{eqnarray}
D_{\alpha\mu} = \frac{1}{2}\left ( \partial_\alpha u_\mu + \partial_\mu u_\alpha \right), \: W_{\alpha\mu} = \frac{1}{2}\left ( \partial_\mu u_\alpha - \partial_\alpha u_\mu \right).
\end{eqnarray}
The elastic stress tensor is expressed as:
\begin{eqnarray}
\sigma_{\alpha\beta}^{e} = -\partial_\alpha n_\gamma \frac{\delta E}{\delta (\partial_\beta n_\gamma)},
\end{eqnarray}
where $E$ represents the Frank-Oseen free energy:
\begin{eqnarray}
E =&& \int dV \biggl ( \frac{K_{11}}{2} (\nabla \cdot \nvec)^2 + \frac{K_{22}}{2} \left ( \nvec\cdot [\nabla \times \nvec] + q_0 \right)^2 \\ && + \frac{K_{33}}{2}[ \nvec\times [\nabla \times\nvec]]^2 \biggr) .
\label{free-energy-eq}
\end{eqnarray}
$K_{11}$, $K_{22}$, $K_{33}$ are the Frank elastic constants, and $q_0=2\pi/p$, with $p$ being the cholesteric pitch. To drive the flow, we used an external body force $\mathbf{g}$, which is added to the r.h.s of Eq.~\eqref{NS-eq}. 

The second set of equations describes the evolution of the director field:
\begin{align}
& \partial_t n_\mu = \frac{1}{\gamma_1} h_\mu -  \frac{\gamma_2}{\gamma_1} n_\alpha D_{\alpha\mu} - u_\gamma \partial_\gamma n_\mu + W_{\mu\gamma}n_\gamma ,
\label{director-time-eq} 
\end{align}
where $\gamma_1=\alpha_3-\alpha_2$ is the rotational viscosity determining the relaxation rate of the director, and $\gamma_2 =\alpha_3+\alpha_2$ is the torsion coefficient characterizing the contribution to the viscous torque from velocity field gradients. The ratio $\gamma_2/\gamma_1$ is known as the aligning parameter, where $\vert \gamma_2/\gamma_1 \vert>1$ indicates flow-aligning systems and $\vert \gamma_2/\gamma_1 \vert<1$ indicates flow-tumbling systems. The molecular field is given by
\begin{eqnarray}
h_\mu = -\frac{\delta E}{\delta n_\mu}.
\end{eqnarray}

The simulations employed a hybrid numerical method. The velocity field was resolved using a standard lattice Boltzmann method (D3Q19 lattice, BGK collision operator, Guo forcing scheme and half-way bounce back boundary conditions)~\cite{kruger2016lattice,succi2018lattice}, with the elastic and viscous stress tensors (except the term proportional to $\alpha_4$) introduced as a force term. The director field equation, Eq.\eqref{director-time-eq}, was solved using a predictor-corrector finite-difference algorithm. On solid boundaries, infinite homeotropic anchoring and no-slip conditions were applied using the bounce-back condition~\cite{kruger2016lattice}. 

The simulations started with the liquid at rest, and the directors primarily aligned perpendicularly to the plates except near the toron, whose configuration was obtained by minimizing the free energy starting from the Ansatz of Ref.~\cite{Coelho_2021}. The material parameters were chosen to be close to those of MBBA at 22$^\circ$C ~\cite{PhysRevE.89.032508}, except for the absolute viscosity (or equivalently, $\alpha_4$), which was doubled to ensure reasonable simulation times. The code was parallelized in CUDA-C, and the simulations were executed on GPUs~\cite{10.5555/1841511}. A typical simulation performance is $\sim$328 MLUPS (Mega Lattice Updates Per Seconds) in a Nvidia Tesla V100 using double precision calculations.

\begin{table}
\caption{\label{tab1} Parameters used in the simulation and physical units.}
\footnotesize
\begin{tabular}{|p{0.28\linewidth}|p{0.28\linewidth}|p{0.28\linewidth}|}
\hline
Symbol&Sim. units & Physical units\\
\hline
$\rho$&1&1088 Kg/m$^{3}$\\
\hline
$\Delta x$&1 & 0.625 $\mu$m\\
\hline
$\Delta t$&1 & 2$\times10^{-9}$ s\\
\hline
$K_{11}$&$1.67 \times 10^{-7}$& $6.4\times 10^{-12}$ N \\
\hline
$K_{22}$&$7.88 \times 10^{-8}$& $3.0\times 10^{-12}$ N \\
\hline
$K_{33}$&$2.62 \times 10^{-7}$& $9.98\times 10^{-12}$ N \\
\hline
$\alpha_1$& 0.0373 & 0.0036 Pa.s\\
\hline
$\alpha_2$& -0.4496 & -0.044 Pa.s\\
\hline
$\alpha_3$& -0.0203 & -0.0020 Pa.s\\
\hline
$\alpha_4$& 0.9318 & 0.091 Pa.s\\
\hline
$\alpha_5$& 0.3084 & 0.030 Pa.s\\
\hline
$\alpha_6$& -0.1617 & -0.016 Pa.s\\
\hline
$P$ & 14 & 8.75 $\mu$m \\
\hline
$L_x$, $L_y$, $L_z$& 56, 56, 16 & 35, 35, 10 $\mu$m\\
\hline
\end{tabular}\\
\end{table}
\normalsize

\section*{Acknowledgements}
We acknowledge financial support from the Portuguese Foundation for Science and Technology (FCT) under the contracts: PTDC/FISMAC/5689/2020 (DOI 10.54499/PTDC/FIS-MAC/5689/2020), UIDB/00618/2020 (DOI 10.54499/UIDB/00618/2020), UIDP/00618/2020 (DOI 10.54499/UIDP/00618/2020), DL 57/2016/CP1479/CT0057 (DOI 10.54499/DL57/2016/CP1479/CT0057) and 2023.10412.CPCA.A2. I.I.S. acknowledges hospitality of the International Institute for Sustainability with Knotted Chiral Meta Matter (WPI-SKCM2) at Hiroshima University, where he was partly working on this article. The experimental research was partly supported by the U.S. Department of Energy, Office of Basic Energy Sciences, Division of Materials Sciences and Engineering, under contract DE-SC0019293 with the University of Colorado at Boulder.

\bibliography{biblio}

\end{document}